# A MINIATURIZED PROGRAMMABLE MULTI-FLUIDIC PNEUMATIC SYSTEM FOR PRECISE CONTROLS OF SAMPLE PREPARATION ENVIRONMENT

Sankar Raju Narayanasamy[1,2], Ramakrishna Vasireddi[3] and Hoi-Ying Holman[1,2]

1. *Berkeley Synchrotron Infrared Structural Biology Imaging Program, Lawrence Berkeley National Laboratory, Berkeley, USA;* 2. *Molecular Biophysics and Integrated Bioimaging Division, Lawrence Berkeley National Laboratory, Berkeley, USA;* 3. *Proxima-1 & Microfluidics Laboratory, Synchrotron SOLEIL, Saint-Aubin, France*

## ABSTRACT

High-density microfluidics is becoming an important experimental platform for studying complex biological systems such as synthetic gene regulatory networks, molecular biocomputating of engineered cells, distributing rapid point-of-care diagnosis, and monitoring pathological environment. Imaging transient bio-chemical reactions happening in these systems at a single particle or cellular level requires precise time-dependent control of sample reaction and imaging conditions at the desired fluidic momentum. In this study, we showed our novel miniaturized and programmable electronic-based pneumatic system to meet the requirement. We demonstrated its capability to control reaction parameters such as concentrations and injection rates in a liposome production system.

**KEYWORDS:** Microfluidics, miniaturized pneumatic system, sample preparation, sample introduction, liposomes

## INTRODUCTION

The intricacy of biological processes can be understood by observing the transient bio-chemical reactions as they happen using high-density microfluidic chips-based imaging studies[1,2]. Very large scale integration of microfluidic channels such as those in high-density microfluidic chips enables researchers to study these reactions. Existing commercially available sample injection methods are based on syringe pumps or pneumatic pressure control pumps which employ lengthy tubing and can be costly. Presently, researchers at imaging user facilities use large power-electrical based pneumatic systems[4] to precisely introduce samples/materials into the integrated microfluidic devices. As the field of biological and biomedical sciences advances, there is an emerging need for more portable and affordable pneumatic systems to accurately introduce complex samples into high-density microfluidic reactors while satisfying the compact space requirement. To meet this need, we develop a miniaturized versatile electronic-based pneumatic system that can be easily adapted to project-specific sample preparation and introduction. This portable system can be scaled up in multiples of 4 miniature pneumatic pumps. Furthermore, it is only a fraction of the total cost of existing commercially available sample injection apparatus.

## EXPERIMENTAL

Our palm-size multi-fluidic pneumatic sample injection device with high-precision sample handling is shown in Figure 1. In the demonstrated liposome synthesis experiment utilizing this new miniaturized pneumatic pressure control system, we used ambient filtered air as pressure feed for the diaphragm pumps. The air can be replaced with specific suitable gas as per biological sample compatibility if needed. The flow rates of each sample introduced into the channels is controlled independently.

## INSTRUMENTATION & RESULTS

The miniature pneumatic pumps are connected to the compact multi-board, which in turn is coupled to the high-drivers and Raspberry Wifi header to meet all inter-connectivity needs. Multiple multi-boards can be easily stacked up in a custom built 3D printed holder. The droplet-stable giant unilamellar vesicles synthesizing microfluidic device[3] was fabricated using maskless lithography procedure, utilizing Dilase-250 table-top high-resolution laser lithography system. We have demonstrated the utilization of the miniaturized programmable multi-throughput multi-fluidic pneumatic system for the synthesis of cell-like liposomes such as droplet-stable giant unilamellar vesicles (dsGUV) in the 10-100 µm range (Figure 2).

## CONCLUSION

The high-precision multi-fluidic bio-sample preparation and injection system can be easily installed in any bioimaging microscopic setups satisfying the compact space requirement in many imaging facilities. Thus this setup is a crucial advance for complex synthetic biology studies. Furthermore, the setup can be extended to multi-level sample injections in single particle and cellular cryo-electron microscopy(cryo-EM) sample preparation systems.

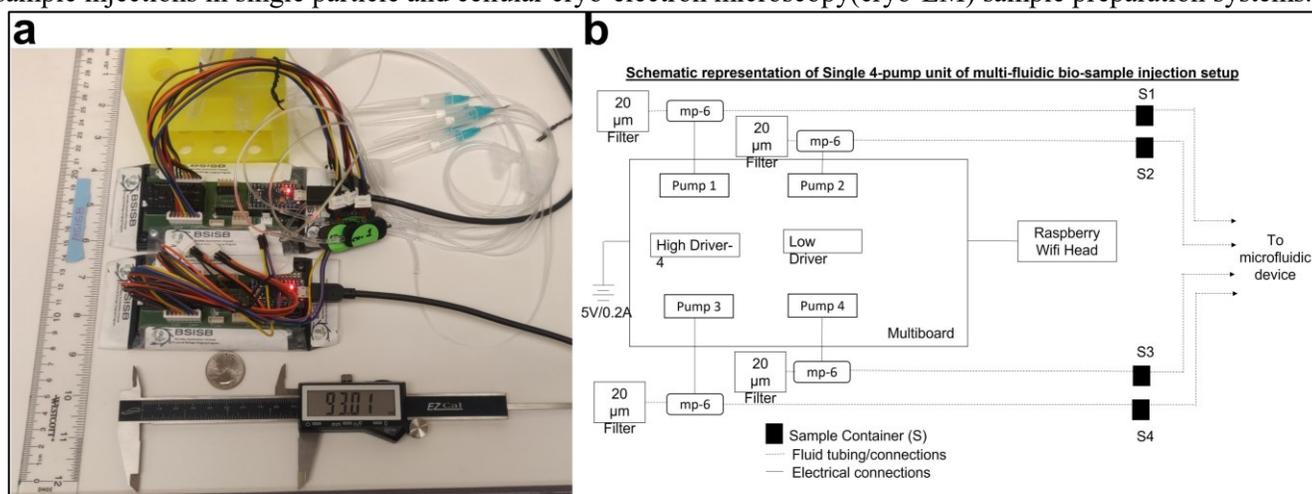

***Figure 1.*** *(**a**) An experimental setup of 2 single 4-pumps units. (**b**) Schematic representation of a single 4-pumps unit. This unit can be easily scaled up as per the number of samples to be controlled and the flow rates for each pump is controlled via PC/Laptop using multi-board application suite.*

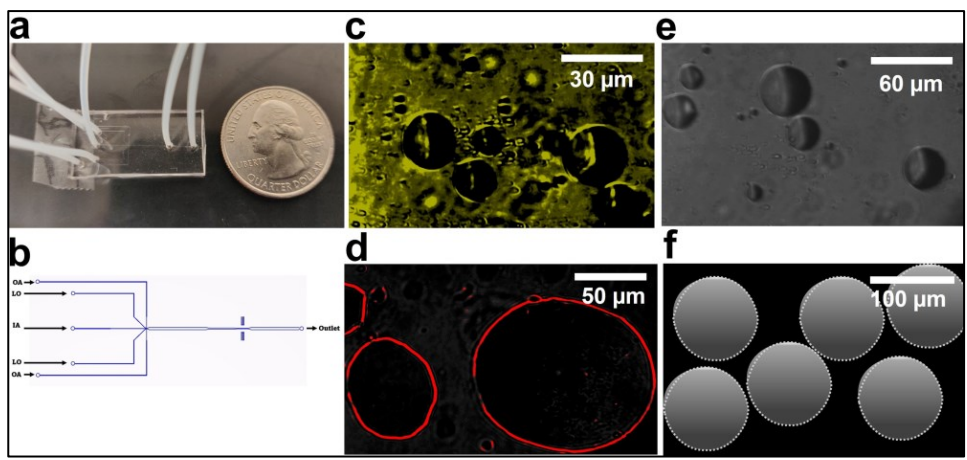

***Figure 2.*** *(**a**) Octanol assisted liposome (OLA) assembly based microfluidic device fabricated using maskless lithography for synthesizing cell-scale micro- liposomes. (**b**) Schematic of the computer aided design of OLA microfluidic device. (**c-d**) Manually produced dsGUVs in the range of 5-25 μm and 30-110 μm. (**e**) Microfluidic produced 30-60 μm dsGUVs, and (**f**)the envisioned 100-μm dsGUVs to be produced.*

**ACKNOWLEDGEMENTS**

This work was conducted through the Berkeley Synchrotron Infrared Structural Biology (BSIS-B) Imaging program, supported by DOE Office of Biological and Environmental Research, under contract no. DE-AC02-05CH11231. The authors also would like to acknowledge the technical team of Bartels Mikrotechnik GmBH, Germany for providing the certain components and specifications to establish this multi-fluidic bio-sample injection setup.

**CONTACT**
   * Hoi-Ying N. Holman; phone: +1-510-486-5943; hyholman@lbl.gov